\documentclass[showpacs,amssymb,floatfix]{revtex4}
\usepackage{graphicx}
\usepackage{dcolumn}
\usepackage{bm}

\begin{document}
\date{\today}
\title{Electromagnetic analog of Rashba spin-orbit interaction in
wave guides filled with ferrite}
\author{Evgeny N.Bulgakov$^1,2$ and Almas F. Sadreev$^{1,2,3}$}
\address{1) Kirensky Institute of Physics, 660036, Krasnoyarsk, Russia\\
2) Krasnoyarsk Pedagogical University,\\ 3) Department of Physics
and Measurement Technology, Link\"{o}ping University, S-581 83
Link\"{o}ping, Sweden}
\begin{abstract} We consider infinitely long electromagnetic wave guide
filled with a ferrite. The wave guide has arbitrary but constant
cross-section $D$. We show that Maxwell equations are equivalent
to the Schr\"odinger equation for single electron in the
two-dimensional quantum dot of the form $D$ with account of the
Rashba spin-orbit interaction. The spin-orbit constant is
determining by components of magnetic permeability of the ferrite.
The upper component of electron spinor function corresponds to the
z-th component electric field, while the down component $\chi$
related to the z-th component of magnetic field by relation
(\ref{magfield}).
\end{abstract}
\maketitle
\section{Introduction}
There is a complete equivalence of the two-dimensional
Schr\"odinger equation for a particle in a hard wall box to
microwave billiards \cite{stockmann}. A wave function is exactly
corresponds to the the electric field component of the TM mode of
electromagnetic field: $\psi(x,y)\leftrightarrow E_z(x,y)$ with
the same Dirichlet boundary conditions. This equivalence is turned
out very fruitful and test a mass of predictions found in the
quantum mechanics of billiards \cite{stockmann}.  In open systems
the probability current density corresponds to the Poynting
vector. The last equivalence allowed to test in particular
universal current statistics in chaotic billiards
\cite{barth,saichev}. Moreover if the resonator is non
homogeneously filled with ferrite, a similarity to quantum
billiards with broken time-reversal symmetry appears
\cite{stockmann}.

In present work we develop this idea of resonators filled with
ferrite to show an equivalence of the Maxwell equations to
electron in quantum dot (QD) with the Rashba spin-orbit
interaction (SOI) \cite{rashba}. This interaction is relevant in a
two-dimensional electron gas, such as is formed in GaAs
heterostructures. While spin-orbit scattering in metals is largely
due to scattering from the metal ions or from impurities, in a
GaAs heterostructure, spin-orbit effects mainly arise from the
asymmetry of the potential creating the quantum well. The
hamiltonian of single electron in the quantum dot with account of
the Rashba SOI has the following form \cite{aronov,morpugo}
\begin{equation}
\label{HSOI} H=\frac{1}{2m^{*}}({\bf p}-\frac{e}{c}{\bf A})^2+
\hbar K [{\bf \sigma\times (p}-\frac{e}{c}{\bf A})]_z
\end{equation}
where $m^{*}=0.023m$ is the effective mass, $\hbar^2 K = 6\times
10^{-10}eV\cdot cm$ is the SOI coefficient in a InAs structures
\cite{aronov}. The dominant mechanism for the SOI in
two-dimensional electron gas (2DEG) is attributed by the Rashba
term \cite{rashba,nitta}. Using the characteristic scale of QD $R$
we rewrite (\ref{HSOI}) in dimensionless form for ${\bf B}=0$
\begin{equation}
\label{hsoi} \widetilde{H}=\left(\matrix{-\nabla^2 &
\beta(\frac{\partial }{\partial{x}}+i\frac{\partial
}{\partial{y}})\cr \beta(-\frac{\partial
}{\partial{x}}+i\frac{\partial }{\partial{y}})&
-\nabla^2\cr}\right),
\end{equation}
where $\beta = 2m^{*}KR$. We consider that the electric field is
directed normal to the plane of QD. If to introduce complex
variables $\xi=x+iy$ and the Cauchy derivative
$\frac{\partial}{\partial \xi}
=\frac{1}{2}\left(\frac{\partial}{\partial
x}-i\frac{\partial}{\partial y}\right)$ the the Schr\"odinger
equation with the Hamiltonian (\ref{hsoi}) takes the following
form
\begin{eqnarray}
\label{zz} \left(\frac{\partial^2}{\partial \xi \partial \xi^{*}}+
\frac{1}{4}\epsilon\right)\phi + \frac{\beta}{2}
\frac{\partial{\chi}}{\partial \xi}=0,\nonumber\\
\left(\frac{\partial^2}{\partial \xi \partial
\xi^{*}}+\frac{1}{4}\epsilon\right)\chi - \frac{\beta}{2}
\frac{\partial{\phi}}{\partial {\xi^{*}}}=0,
\end{eqnarray}
where $\phi, \chi$ are the components of the spin state and
$\epsilon$ is the eigen energy.
\section{Basic electromagnetic equations}
Let us consider electromagnetic waves in a wave guide filled with
a ferrite with a magnetization $M_0$ directed along the z axis and
with the following magnetic permutability
\begin{eqnarray}
\label{chi} \hat\mu=1+4\pi\hat\chi,\nonumber\\
\hat\chi=\left(\matrix{\chi_{xx} &  \chi_{xy} &  0\cr
                           -\chi_{xy} & \chi_{xx} & 0\cr
                           0 &  0  & \chi_{zz}\cr}\right).
\end{eqnarray}
By relation ${\bf B}=\hat{\mu}{\bf H}$ the Maxwell equations take
the following form
\begin{eqnarray}
\label{ME1} \left(\frac{\partial{E_z}}{\partial
y}-\frac{\partial{E_y}}{\partial
z}\right)=-ik(\mu_{xx}H_x+\mu_{xy}H_y),\nonumber\\
\left(\frac{\partial{E_x}}{\partial
z}-\frac{\partial{E_z}}{\partial
x}\right)=-ik(\mu_{yx}H_x+\mu_{yy}H_y),\\
\left(\frac{\partial{E_y}}{\partial
x}-\frac{\partial{E_x}}{\partial
y}\right)=-ik\mu_{zz}H_z;\nonumber
\end{eqnarray}
\begin{eqnarray}
\label{ME2} \left(\frac{\partial{H_z}}{\partial
y}-\frac{\partial{H_y}}{\partial z}\right)=ikE_x,\nonumber\\
\left(\frac{\partial{H_x}}{\partial
z}-\frac{\partial{H_z}}{\partial x}\right)=ikE_y,\\
\left(\frac{\partial{H_y}}{\partial
x}-\frac{\partial{H_x}}{\partial y}\right)=ikE_z,\nonumber
\end{eqnarray}
where frequency of electromagnetic waves $\omega=ck$. The first
couple of the Maxwell equations $\nabla {\bf E}=0, \nabla {\bf
B}=0$ follows from the second couple of equations (\ref{ME1}) and
(\ref{ME2}).

\begin{figure}[t]
\includegraphics[width=.3\textwidth]{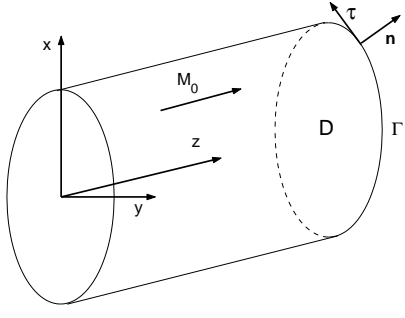}
\caption{Schematical view of electromagnetic wave guide with
constant cross-section} \label{fig1}
\end{figure}

Let us consider the wave guide infinitely long in $z$ direction.
We take that a cross-section of the wave guide is constant in $x,
y$ plane as shown in Fig. \ref{fig1}.
Such a geometry of the waveguide allows to separate variables $z$
and $x, y$ and write
\begin{equation}\label{separation}
  {\bf E}(x,y,z)=e^{ik_z z}{\bf E}(x,y),\quad
  {\bf H}(x,y,z)=e^{ik_z z}{\bf H}(x,y).
\end{equation}
Then Eqs \ref{ME2} reduce to
\begin{equation}
\label{ME3} \frac{\partial}{\partial
y}\left(\frac{\partial{H_z}}{\partial
y}-\frac{\partial{H_y}}{\partial
z}\right)-\frac{\partial}{\partial
x}\left(\frac{\partial{H_x}}{\partial
z}-\frac{\partial{H_z}}{\partial
x}\right)=ik\frac{\partial{E_x}}{\partial
y}-ik\frac{\partial{E_y}}{\partial x}.
\end{equation}
Combining this equation with the third equation of (\ref{ME1}) we
obtain from (\ref{ME3})
\begin{equation}\label{ME4}
\nabla_{\bot}^2 H_z-\frac{\partial}{\partial z}
\left(\frac{\partial{H_x}}{\partial x}
+\frac{\partial{H_y}}{\partial y}\right) +k^2\mu_{zz}H_z=0,
\end{equation}
where $\nabla_{\bot}=\frac{\partial^2}{\partial
x^2}+\frac{\partial^2}{\partial y^2}$ is the two-dimensional
Laplace operator.

The next task is to exclude $H_y$ and $H_x$ from (\ref{ME4}). Eq.
$\nabla {\bf B}=0$ gives
\begin{equation}\label{ME5}
\mu_{xx}\left(\frac{\partial{H_x}}{\partial x}
+\frac{\partial{H_y}}{\partial
y}\right)+\mu_{xy}\left(\frac{\partial{H_y}}{\partial x}
-\frac{\partial{H_x}}{\partial
y}\right)+\mu_{zz}\frac{\partial{H_z}}{\partial z}=0.
\end{equation}
Substituting the third equation (\ref{ME2}) into (\ref{ME5}) we
obtain $$-\left(\frac{\partial{H_x}}{\partial x}
+\frac{\partial{H_y}}{\partial y}\right)
=\frac{\mu_{zz}}{\mu_{xx}}\frac{\partial{H_z}}{\partial z}
+ik\frac{\mu_{xy}}{\mu_{xx}}E_z.$$ Substituting this equation into
(\ref{ME4}) we obtain finally the first equation for $E_z$ and
$H_z$:
\begin{equation}\label{Ez}
\nabla_{\bot}^2H_z(x,y)+\left\{-k_z^2\frac{\mu_{zz}}{\mu_{xx}}
+k^2\mu_{zz}\right\}H_z(x,y)-kk_z\frac{\mu_{xy}}{\mu_{xx}}E_z(x,y)=0.
\end{equation}

Let us derive the next equation for $E_z$ and $H_z$. The first
couple of equations (\ref{ME1}) gives us:
\begin{equation}\label{ME6}
\frac{\partial}{\partial y}\left(\frac{\partial{E_z}}{\partial
y}-\frac{\partial{E_y}}{\partial z}\right)
-\frac{\partial}{\partial x}\left(\frac{\partial{E_x}}{\partial z}
-\frac{\partial{E_z}}{\partial x}\right)=
-ik\frac{\partial}{\partial y}(\mu_{xx}H_x+\mu_{xy}H_y)
+ik\frac{\partial}{\partial x}(\mu_{yx}H_x+\mu_{yy}H_y).
\end{equation}
Because of Eqs (\ref{chi}), (\ref{ME2}) and (\ref{separation})
this equation can be written as follows
\begin{equation}\label{ME7}
(\nabla_{\bot}^2+\mu_{xx}k^2-k_z^2)E_z=ik\mu_{yx}\left(\frac{\partial{H_x}}{\partial
x} +\frac{\partial{H_y}}{\partial y}\right).
\end{equation}
On the other hand, Eq. $\nabla {\bf B}=0$ gives $$
\mu_{xx}\left(\frac{\partial{H_x}}{\partial x}
+\frac{\partial{H_y}}{\partial
y}\right)+ik\mu_{xy}E_z+\mu_{zz}\frac{\partial{H_z}}{\partial
z}=0.$$  As a result we obtain finally from (\ref{ME2}) and
(\ref{ME7}) the second equation for $E_z$ and $H_z$:
\begin{equation}\label{Hz}
\left\{\nabla_{\bot}^2+(\mu_{xx}k^2-k_z^2)+k^2\frac{\mu_{xy}}{\mu_{xx}}\right\}E_z
+kk_z\frac{\mu_{xy}}{\mu_{xx}}H_z=0.
\end{equation}
\section{Boundary conditions}
Let us consider the boundary conditions for these relevant
electromagnetic fields $E_z$ and $H_z$.  Assume that surface of
the wave guide shown in Fig. \ref{fig1} is perfect metal. The
boundary conditions are the following
\begin{equation}\label{boundary}
  E_z(x,y)\vert_{_{\Gamma}}=0,\quad E_{\tau}(x,y)\vert_{_{\Gamma}}=0
\end{equation}
where $\tau$ is the unit vector directed along the boundary
$\Gamma$ in the wave guide cross-section. Our task is to express
$E_{\tau}$ through the relevant fields $E_z$ and $H_z$. Write the
Maxwell equations (\ref{ME1}), (\ref{ME2}) and (\ref{separation})
as follows
\begin{eqnarray} \label{BC1} \frac{\partial{E_z}}{\partial
y} -ik_zE_y=-ik(\mu_{xx}H_x+\mu_{xy}H_y),\nonumber\\
ik_zE_x-\frac{\partial{E_z}}{\partial x}
=-ik(\mu_{yx}H_x+\mu_{yy}H_y)
\end{eqnarray}
and
\begin{eqnarray}
\label{BC2} ik_zH_y=\frac{\partial{H_z}}{\partial y}
-ikE_x,\nonumber\\
ik_zH_x=\frac{\partial{H_z}}{\partial x}
+ikE_y.
\end{eqnarray}
From these equations we can express fields $E_x$ and $E_y$ via the
fields $E_z$ and $H_z$. A tedious but straightforward algebra
gives the following equations:
$$E_x=\frac{1}{D}\Bigg\{\left(\frac{\partial{H_z}}{\partial x}
\right)[-ik^3\mu_{xy}\mu_{xx}-ik\mu_{xy}(k_z^2-k^2\mu_{xx})]$$ $$
+\left(\frac{\partial{H_z}}{\partial y} \right)
[-ik^3\mu_{xy}^2+ik\mu_{xx}(k_z^2-k^2\mu_{xx})]
-ik_zk^2\mu_{xy}\left(\frac{\partial{E_z}}{\partial y}\right)
-ik_z(k_z^2-k^2\mu_{xx})\left(\frac{\partial{E_z}}{\partial x}
\right)\Bigg\},$$
$$E_y=\frac{1}{D}\Bigg\{\left(\frac{\partial{H_z}}{\partial x}
\right)[+ik^3\mu_{xy}^2-ik\mu_{xx}(k_z^2-k^2\mu_{xx})]$$
\begin{equation}\label{ExEy}
+\left(\frac{\partial{H_z}}{\partial y} \right)
[-ik^3\mu_{xx}\mu_{xy}-ik\mu_{xy}(k_z^2-k^2\mu_{xx})]
+ik_zk^2\mu_{xy}\left(\frac{\partial{E_z}}{\partial x}\right)
-ik_z(k_z^2-k^2\mu_{xx})\left(\frac{\partial{E_z}}{\partial y}
\right)\Bigg\},
\end{equation}
where $D=k^4\mu_{xy}^2+(k_z^2-k^2\mu_{xx})^2$.

Since $\overrightarrow{\tau}=(-n_y n_x)$, the second boundary
condition (\ref{boundary}) takes the following form
$-n_yE_x+n_xE_y\vert_{_\Gamma}=0$. Taking into account the first
boundary condition (\ref{boundary}) we obtain finally from
(\ref{ExEy})
\begin{eqnarray}\label{boundary1}
\mu_{xy}\left\{kk_z\left(\frac{\partial{E_z}}{\partial n}
\right)-k_z^2(\overrightarrow{\tau}\nabla_{\bot}H_z)\right\} +
\left(\frac{\partial{H_z}}{\partial
n}\right)[k^2(\mu_{xy}^2+\mu_{xx}^2)-k_z^2\mu_{xx}]\vert_{_\Gamma}=0\nonumber\\
E_z\vert_{_\Gamma}=0.
\end{eqnarray}
\section{SOI equivalent microwave equations}
In order to write Eqs (\ref{Ez}) and (\ref{Hz}) in form equivalent
to the Schr\"odinger equation (\ref{zz}) with the Rashba SOI let
us introduce the following notations
\begin{eqnarray}\label{notations}
h_z=iH_z,\nonumber\\
\epsilon_1=k^2-\frac{k_z^2}{\mu_{xx}},\nonumber\\
\epsilon_2=k^2\mu_{xx}-k_z^2+\frac{k^2\mu_{xy}^2}{\mu_{xx}},\nonumber\\
\nu=kk_z\kappa,\quad i\kappa=\frac{\mu_{xy}}{\mu_{xx}},
\end{eqnarray}
where \cite{lax}
\begin{eqnarray}
\label{mu} \mu_{xx}(\omega)=1+\frac{4\pi M_0\omega_0 \Omega
}{\Omega^2-\omega^2},\quad \mu_{xy}(\omega)=i\frac{4\pi
M_0\omega_0 \omega}{\Omega^2-\omega^2},\\ \omega_0=\gamma
(H_0-4\pi M_0), \nonumber\\ \Omega^2=\omega_0 (\omega_0+\gamma
H_a),\nonumber\\ H_a=8\pi K_a/M.\nonumber
\end{eqnarray}
Here $H_0$ is external constant magnetic field applied along the
z-axis, i.e. along the direction of magnetization $M_0$ and $H_a$
is effective anisotropy field directed along the x-axis.

Let us introduce an auxiliary function $u_2(x,y)$:
\begin{equation}\label{u2}
  h_z=\lambda\frac{\partial{u_2}}{\partial{\xi}},
\end{equation}
where $\xi=x+iy$, and $\lambda$ is some coefficient which will be
determined below . Then Eqs (\ref{Ez}) and (\ref{Hz}) take the
following form
\begin{eqnarray}\label{eq1}
\left(4\frac{\partial^2}{\partial \xi \partial \xi^{*}}+
\epsilon_2\right)E_z+\lambda\nu\frac{\partial{u_2}}{\partial{\xi}}=0,\nonumber\\
\left(4\frac{\partial^2}{\partial \xi \partial \xi^{*}}+
\epsilon_1\right)\lambda\frac{\partial{u_2}}{\partial{\xi}}+\nu
E_z =0.
\end{eqnarray}
The first equation (\ref{eq1}) can be written as
\begin{equation}\label{eq2}
  E_z=-\frac{1}{\epsilon_2}\left(4\frac{\partial^2}{\partial \xi \partial \xi^{*}}+
\lambda\nu\frac{\partial{u_2}}{\partial{\xi}}\right).
\end{equation}
Let us choose $\lambda^2=4/\epsilon_2$. Then the second equation
(\ref{eq1}) takes the following form
\begin{equation}\label{eq3}
\frac{\partial}{\partial{\xi}}\left\{\Bigg[4\frac{\partial^2}{\partial
\xi \partial
\xi^{*}}+\left(\epsilon_1-\frac{\nu^2}{\epsilon_2}\right)\Bigg]u_2-
\frac{2\nu}{\sqrt{\epsilon_2}}\frac{\partial{E_z}}{\partial{\xi^{*}}}\right\}=0.
\end{equation}

The solution of equation $\frac{\partial g}{\partial{\xi}}=0$ is
any function $g=g(\xi^{*})$ to give us the following equation:
\begin{equation}\label{eq4}
\Bigg[4\frac{\partial^2}{\partial \xi \partial
\xi^{*}}+\left(\epsilon_1-\frac{\nu^2}{\epsilon_2}\right)\Bigg]u_2-
\frac{2\nu}{\sqrt{\epsilon_2}}\frac{\partial{E_z}}{\partial{\xi^{*}}}=g(\xi^{*}).
\end{equation}
One can see that the magnetic field (\ref{u2}) is invariant under
gauge transformations $u_2\rightarrow \chi=u_2+f(\xi^{*})$. Let us
make this gauge transformation and substitute into Eq.
(\ref{eq4}). Then we obtain the following equation for transformed
$\chi$:
\begin{equation}\label{eq5}
  4\frac{\partial^2 {\chi}}{\partial \xi \partial
\xi^{*}}+\left(\epsilon_1-\frac{\nu^2}{\epsilon_2}\right)\chi-
\frac{2\nu}{\sqrt{\epsilon_2}}\frac{\partial{E_z}}{\partial{\xi^{*}}}=0,
\end{equation}
provided that we have chosen the gauge as follows:
\begin{equation}\label{gauge}
\left(\epsilon_1-\frac{\nu^2}{\epsilon_2}\right)f(\xi^{*})=g(\xi^{*}).
\end{equation}
Finally combining this equation with the first equation
(\ref{eq1}) we obtain the following system of equations
\begin{eqnarray}\label{SOIEM}
\left(\frac{\partial^2}{\partial \xi \partial \xi^{*}}+
\frac{1}{4}\epsilon_2\right)E_z+\frac{\nu}{2\sqrt{\epsilon_2}}
\frac{\partial{\chi}}{\partial{\xi}}=0,\nonumber\\
\left(\frac{\partial^2 }{\partial \xi \partial
\xi^{*}}+\frac{1}{4}\left(\epsilon_1-\frac{\nu^2}{\epsilon_2}\right)\right)\chi-
\frac{\nu}{2\sqrt{\epsilon_2}}\frac{\partial{E_z}}{\partial{\xi^{*}}}=0.
\end{eqnarray}

If to compare these equations with the Schr\"odinger equation for
electron in quantum dot with SOI (\ref{zz}) one can see that the
electric field $E_z$ is equivalent to the upper spinor component
$\phi$ while $\chi$ is down component which related to magnetic
field by relation
\begin{equation}\label{magfield}
  H_z=-\frac{2i}{\sqrt{\epsilon_2}}\frac{\partial{\chi}}{\partial{\xi}}.
\end{equation}
The parameter $\frac{\nu}{\sqrt{\epsilon_2}}$ is completely
equivalent to the SOI constant $\beta$.  If to substitute
relations (\ref{notations}) one can see that Eq. (\ref{SOIEM}) is
the equation for eigen functions $E_z$ and $\chi$ and eigenvalues
$k^2$. The eigenvalues of Eq. (\ref{SOIEM}) will coincide with the
eigenenergies of Eq. (\ref{zz}) if to choose by variation of
magnetic permutability $\mu_{xx}, \mu_{xy}$ and $k_z$
\begin{equation}\label{eigenvalue}
  \epsilon_2=\epsilon_1-\frac{\nu^2}{\epsilon_2},
\end{equation}
we can obtain exact correspondence between Eq. (\ref{zz}) and
(\ref{SOIEM}). Only difference in boundary conditions is
remaining.

This work has been partially supported by Russian Foundation for
Basic research Grant 03-02-17039.

\end{document}